\documentclass[twocolumn,prl,epsf,graphics,psfig,superscriptaddress]{revtex4}

\usepackage{graphicx,graphics} % Include figure files
\usepackage[dvips,bookmarks=false]{hyperref}
\usepackage{graphicx}
\usepackage{dcolumn}
\usepackage{eucal}
\usepackage[dvips]{epsfig}

\begin{document}
\title{Optical absorption spectrum in disordered semiconductor multilayers}
\author{Alireza Saffarzadeh} \email{a-saffar@tpnu.ac.ir}
\affiliation{Department of Physics, Payame Noor University,
Nejatollahi Street, 159995-7613 Tehran, Iran}
\affiliation{Computational Physical Sciences Laboratory,
Department of Nano-Science, Institute for Research in Fundamental
Sciences (IPM), P.O. Box 19395-5531, Tehran, Iran}
\author{Leili Gharaee} \affiliation{Department of Physics, Payame Noor University,
Nejatollahi Street, 159995-7613 Tehran, Iran}
\date{\today}

\begin{abstract}
The effects of chemical disorder on the electronic and optical
properties of semiconductor alloy multilayers are studied based on
the tight-binding theory and single-site coherent potential
approximation. Due to the quantum confinement of the system, the
electronic spectrum breaks into a set of subbands and the
electronic density of states and hence the optical absorption
spectrum become layer-dependent. We find that, the values of
absorption depend on the alloy concentration, the strength of
disorder, and the layer number. The absorption spectrum in all
layers is broadened because of the influence of disorder and in
the case of strong disorder regime, two optical absorption bands
appear. In the process of absorption, most of the photon energy is
absorbed by the interior layers of the system. The results may be
useful for the development of optoelectronic nanodevices.
\end{abstract}
\maketitle
%{\bf PACS.}

\section{Introduction}
Low-dimensional quantum structures, such as quantum wells and
superlattices, wires and dots have attracted great interest in the
last few years, due to their possible application in nanodevices
\cite{Harrison}. Among these structures, multilayers and
superlattices, with thickness of few nanometers, are of
considerable experimental and theoretical interest because of
their specific electronic and optical properties and many
promising areas of applications. For example, multilayer
semiconductors are very useful for laser diodes, leading to low
threshold current, high power, and weak temperature dependence
devices \cite{Razeghi}.

It has been demonstrated that, by using semiconductor
nanostructures instead of bulk structures, the threshold current
may be reduced by more than ten times due to the abrupt energy
dependence of the density of states in low-dimensional structures
which can enhance the light amplification mechanisms and thus
allows lasing to occur at lower currents \cite{Asada}.
Furthermore, the optical absorption in semiconductors which is
governed by the electronic density of states, is strongly affected
by the impurities \cite{Fuchs,Shinozuka1,Shinozuka2}. For
instance, using a semiconducting alloy A$_{1-x}$B$_x$, composed of
two semiconductors A and B with different band gap energies, one
can tune the operation of an optical device at desired wavelength.

There are many different theoretical approaches that can be used
for studying the electronic properties of disordered
semiconductors (see for instance \cite{Gonis}). Among them, the
coherent potential approximation (CPA) \cite{Soven,Velick,Gonis}
is one of the most widely used methods to study chemically
disordered systems. In this approximation the multiple scattering
on a single site is taken into consideration and this approach is
fairly good for any values of bandwidth and scattering potential.

According to the work of Onodera and Toyozawa
\cite{Onodera,Toyozawa}, the optical absorption spectra of
substitutional binary solid solutions can be classified into
persistence and amalgamation types in a unified theory. In the
persistence type, two exciton peaks corresponding to the two
constituent substances are observed, as seen in alkali halides
with mixed halides such as KCl$_{1-x}$Br$_x$. In the amalgamation
type, however, only one exciton state appears, which can be seen
in alkali halides with mixed alkalis such as K$_x$Rb$_{1-x}$Cl.
These two typical cases can be distinguished by a parameter which
is the ratio of the difference of the atomic energies in the two
constituent substances to the energy-band width. In the
persistence type, this difference is large compared to the
bandwidth and the electronic spectrum is split into two bands,
while, when this ratio is small, the impurity band is united with
the host band and in this case, which corresponds to the
amalgamation type, a single band appears. We should note that the
transition from the amalgamation type to the persistence type also
depends on the value of impurity concentration \cite{Onodera}.

Based on the Onodera-Toyozawa theory and the CPA, optical
properties of various disorder systems have been investigated
\cite{Yoshikawa,Shinozuka1,Shinozuka2,Bakalis,Takahashi}.
Recently, Shinozuka \textit{et al.} \cite{Shinozuka1,Shinozuka2}
studied the effect of chemical disorder on the electronic and
optical properties of bulk, quantum well and wire systems within
the CPA. Because of the absence of periodicity in the confinement
directions in the quantum wells and wires, the electronic and
optical properties of these systems become site-dependent. In
their calculations, however, such a dependence has not been
included. This site-dependence may be important in operation of
nanoscale devices and will be included in our calculations.

The aim of this work is to extend the CPA for the doped
semiconductor multilayers to investigate the influence of
substitutional disorder on the local density of states (LDOS) and
optical absorbtion spectrum in each layer of the system and in
both the persistence and amalgamation types. The paper is
organized as follows: in Sec. II, by applying the single-site
approximation to the layered structures in the presence of
chemical disorder, we present our theory and derive equations for
the LDOS and optical absorption. The numerical results and
discussion for the behavior of desired quantities in terms of the
alloy concentration and the scattering-strength parameter are
given in Sec. III. The concluding remarks are given in section IV.

\section{Model and formalism}
We consider a multilayer structure where the layers are stacked
along the $z$-direction and labeled by integer number $n$. The
number of layers in this direction is $N_z$, hence $1\leq n\leq
N_z$. The lattice structure of the system is assumed to be a
simple cubic with lattice spacing $a$ and the (001) orientation of
the layers is taken to be normal to these layers. The multilayer
is a semiconducting alloy of the form $A_{1-x}B_x$, where $A$ atom
(such as Si) and $B$ atom (such as Ge) occupy randomly the lattice
sites with concentration $x$. We use the single-band tight-binding
approximation with nearest-neighbor hopping and the on-site delta
function-like potential. The Hamiltonian for a single electron (or
a Frenkel exciton) in this system is given by
\begin{equation}\label{1}
H=\sum_{\mathbf{r},n}\sum_{\mathbf{r}',n'}[u_{\mathbf{r},n}\delta_{n,n'}\delta_{\mathbf{r},\mathbf{r}'}
-t_{\mathbf{r}n,\mathbf{r}'n'}]\mid \mathbf{r},n\rangle\langle
\mathbf{r}',n'\mid \ ,
\end{equation}
where $\mathbf{r}$ denotes the positional vector in the $x-y$
plane and the on-site potential $u_{\mathbf{r},n}$ is assumed to
be $E_A$ and $E_B$ with probabilities $1-x$ and $x$ when the
lattice site ($\mathbf{r},n$) is occupied by the $A$ and $B$
atoms, respectively. The arrangement of these atoms is completely
random. The hopping integral $t_{\mathbf{r}n,\mathbf{r}'n'}$ is
equal to $t$ for the nearest neighbor sites and zero otherwise. We
assume that the hopping integral only depends on the relative
position of the lattice sites, that is, only the diagonal disorder
is considered. The introduced randomness is treated by the CPA
\cite{Soven,Gonis}. In this approximation, one seeks to replace a
disordered alloy with an effective periodic medium. This CPA
medium is characterized by a local (i.e., momentum-independent),
energy-dependent self-energy, $\Sigma(\omega)$, called the
coherent potential. In some disordered low dimensional systems
such as the quantum wires \cite{Nikolic,Saffar} and the
multilayers \cite{Okiji,Hasegawa}, the self-energy depends on the
position of atomic sites, due to the absence of translational
symmetry in some directions. Therefore, in this study the
self-energy depends on the layer number, i.e.,
$\Sigma(\omega)\equiv\Sigma_n(\omega)$ because of the quantum
confinement along the $z$-direction. Accordingly, the multilayer
Hamiltonian $H$ is replaced by an effective medium Hamiltonian
defined by
\begin{equation}\label{2}
\mathcal{H}_{eff}=\sum_{\mathbf{r},n}\sum_{\mathbf{r}',n'}
[\Sigma_{n}(\omega)\delta_{n,n'}\delta_{\mathbf{r},\mathbf{r}'}-t_{\mathbf{r}n,\mathbf{r}'n'}]\mid\mathbf{r},n\rangle\langle
\mathbf{r}',n'\mid\ .
\end{equation}

\begin{figure}
\centerline{\includegraphics[width=0.85\linewidth]{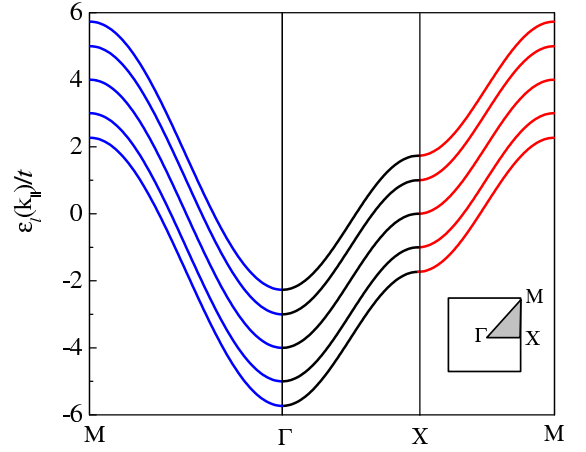}}
\caption{(Color online) Band structure of the clean multilayer
with $N=5$ along the high symmetry axes in the 1BZ (see inset).}
\end{figure}

The physical properties of the real system can be obtained from
the configurationally averaged Green's function $\langle
G\rangle_{\mathrm{av}}=\langle(\omega-H)^{-1}\rangle_{\mathrm{av}}$
which is replaced in the CPA with an effective medium Green's
function $\bar G=(\omega-\mathcal{H}_{eff})^{-1}$ defined by the
following Dyson equation \cite{Saffar}
\begin{eqnarray}\label{3}
\bar{G}_{n,n'}(\mathbf{r},\mathbf{r}';\omega)&=&G^0_{n,n'}(\mathbf{r},\mathbf{r}';\omega)
+\sum_{n''=1}^{N_z}\sum_{\mathbf{r}''}G^0_{n,n''}(\mathbf{r},\mathbf{r}'';\omega)
\nonumber\\
&&\times\Sigma_{n''}(\omega)
\bar{G}_{n'',n'}(\mathbf{r}'',\mathbf{r}';\omega)\ ,
\end{eqnarray}
where $G^0$ is the clean system Green's function and its matrix
element is given by
\begin{equation}\label{4}
G^0_{n,n'}(\mathbf{r},\mathbf{r}';\omega)=\frac{1}{N_\parallel}\sum_{\ell=1}^{N_z}
\sum_{\mathbf{k}_\parallel}G^0_{n,n'}(\ell,\mathbf{k}_\parallel;\omega)
\,e^{i\mathbf{k}_\parallel\cdot(\mathbf{r}-\mathbf{r}')} \  ,
\end{equation}
here, $G^0_{n,n'}(\ell,\mathbf{k}_\parallel;\omega)$ is the mixed
Bloch-Wannier representation of $G^0$ and is expressed as
\begin{equation}\label{GLK}
G^0_{n,n'}(\ell,\mathbf{k}_\parallel;\omega)=\frac{\frac{2}{(N_z+1)}\sin(\frac{\ell\pi}{N_z+1}n)
\sin(\frac{\ell\pi}{N_z+1}n')}{\omega+i\eta-\varepsilon_{\ell}(\mathbf{k}_\parallel)}\
,
\end{equation}
and
\begin{equation}\label{42}
\varepsilon_{\ell}(\mathbf{k}_\parallel)=-2t[\cos(k_xa)+\cos(k_ya)+\cos(\frac{\ell\pi}{N_z+1})]\
,
\end{equation}
is the clean system band structure. In Eq. (\ref{4}),
$\mathbf{k}_\parallel\equiv(k_x,k_y)$ is a wave vector parallel to
the layer and the summation is over all the wave vectors in the
first Brillouin zone (1BZ) of the two-dimensional lattice
\cite{Gonis}, $\ell$ is the mode of the subband, $N_\parallel$ is
the number of lattice sites in each layer, and $\eta$ is a
positive infinitesimal.

Now, to determine the coherent potential in each layer (say $n$),
we consider an arbitrary chosen site in the effective medium
(layer) describe by $\Sigma_n$. Then, we apply the condition that
the average scattering of a carrier by the chosen site in the
medium is zero. This condition for any site $\mathbf{r}$ in the
$n$th layer is given by
\begin{equation}\label{5}
\langle
\tilde{t}_{n,\mathbf{r}}\rangle_\mathrm{av}=\frac{(1-x)[E_A-\Sigma_n]}{1-[E_A-\Sigma_n]F_n}
+\frac{x[E_B-\Sigma_n]}{1-[E_B-\Sigma_n]F_n}=0\ ,
\end{equation}
where, $\tilde{t}_{n,\mathbf{r}}$ is the single-site $t$-matrix
which represents the multiple scattering of carriers by $A$ and
$B$ atoms in the effective medium,
$\langle\cdots\rangle_\mathrm{av}$ denotes average over the
disorder in the system, and
$F_n(\omega)=\bar{G}_{n,n}(\mathbf{r},\mathbf{r};\omega)$ is the
diagonal matrix element of the Green's function of the $n$th
effective layer and for $n=1,\cdots,N_z$ can be written as
\begin{equation}\label{6}
F_n(\omega)=\frac{a^2}{{4\pi^2}}\sum^{N_z}_{\ell=1}\int_{\mathrm{1BZ}}d\mathbf{k}_\parallel\,\bar{G}_{n,n}
(\ell,\mathbf{k}_\parallel;\omega)\  .
\end{equation}
We should note that, the effective Green's function, $\bar G$,
depends on $\ell$ and the two dimensional wave vector
$\mathbf{k}_\parallel$ via $G^0$.

From Eq. (\ref{5}) one can derive an equation for the self-energy
of $n$th layer. Then, using such an equation and also Eq.
(\ref{3}), which gives a system of linear equations, one can
obtain self-consistently the self-energy $\Sigma_n(\omega)$ and
the Green's function $F_n(\omega)$ in each layer. Then, the LDOS
in the $n$th layer is calculated by
\begin{equation}\label{7}
g_n(\omega)=-\frac{1}{\pi}\,\mathrm{Im}\,F_n(\omega)\  .
\end{equation}

In order to calculate the optical absorption spectrum, we assume
that both the $A$ and $B$ atoms have equal transition dipole
moments \cite{Onodera,Toyozawa}. Accordingly, when the $\ell$th
subband is optically excited the layer-dependent optical
absorption is given by the $\mathbf{k}_\parallel=0$ component of
the LDOS. The reason is that $\mathbf{k}_\parallel=0$ (i.e.,
$\Gamma$ point in the inset of Fig. 1) is the bottom of the
conduction band at each layer. We should note that, in the case of
bulk materials, $\mathbf{k}=0$ (in three dimensions) is the
minimum point in the exciton band. In this study
$\varepsilon_\ell(\mathbf{k}_\parallel=0)=-2t[2+\cos(\ell\pi/(N_z+1)]$
gives the minimum energy for each subband. Therefore, the optical
absorption of the $n$th layer, due to the creation of an exciton
in the system, can be expressed as
\begin{equation}\label{optic}
A_n(\omega)=-\frac{1}{\pi}\,\mathrm{Im}\sum_{\ell=1}^{N_z}\bar{G}_{n,n}(\ell,\mathbf{k}_\parallel=0;\omega)\
.
\end{equation}

If it is assumed that, the self-energy does not depend on the
layer number (i.e., the assumption of translational invariant
along the $z$-direction), then a simple expression for Eq.
(\ref{optic}) is obtained, as it was assumed in Refs.
\cite{Shinozuka1,Shinozuka2}. In the present work, the behavior of
this function in each layer of the system which depends on the
number of layers, alloy concentration and the strength of
scattering processes via the self-energy will be studied in Sec.
III.
\begin{figure}
\centerline{\includegraphics[width=1.1\linewidth]{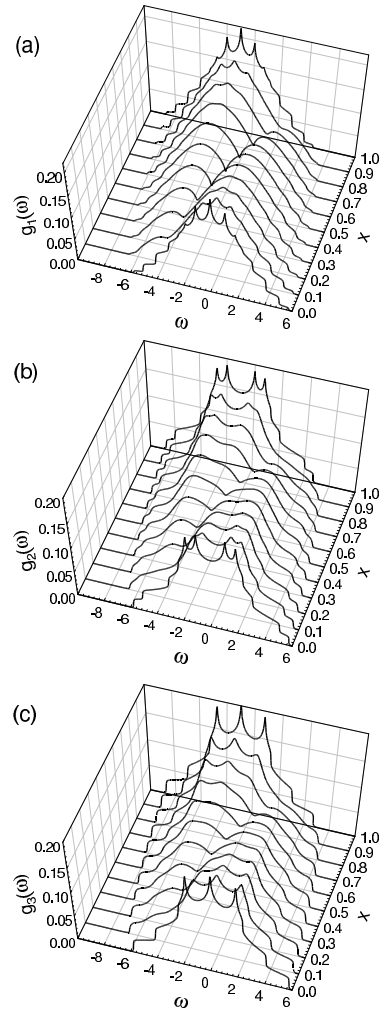}}
\caption{The LDOS for a disordered semiconductor multilayer with
$E_B=-3\,t$ as a function of energy and alloy concentration for
the layers (a) $n=1,5$, (b) $n=2,4$ and (c) $n=3$, respectively.
$\omega$ is measured in units of $t$.}
\end{figure}

\begin{figure}
\centerline{\includegraphics[width=1.1\linewidth]{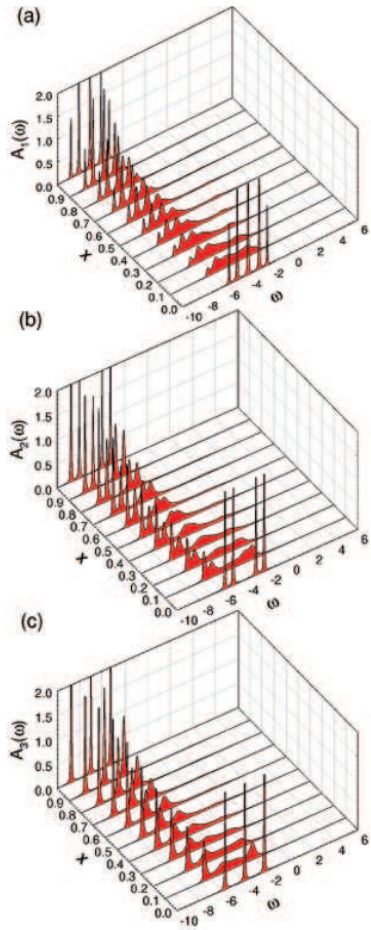}}
\caption{(Color online) The optical absorption spectrum for a
disordered semiconductor multilayer with $E_B=-3\,t$ as a function
of energy and alloy concentration for the layers (a) $n=1,5$, (b)
$n=2,4$ and (c) $n=3$, respectively. $\omega$ is measured in units
of $t$.}
\end{figure}

\section{Results and discussion}
To study the electronic and optical properties of the system, we
perform our numerical results for $N=5$ and characterize the
scattering-strength parameter by $\Delta=\frac{|E_A-E_B|}{W}$,
where $W\simeq 6t$ is the half-bandwidth of either constituent. We
also emphasize that, the calculation of $A_n(\omega)$ is not
restricted to the lowest $\ell=1$ subband, because the results
showed that $\mathbf{k}_\parallel=0$ component in other $\ell$
subbands is finite and hence, all the subband contributions must
be included.

In order to see the effect of quantum confinement on the clean
system, $\varepsilon_{\ell}(\mathbf{k}_\parallel)$ is shown in
Fig. 1 along the high symmetry directions in the 1BZ. We observe
five energy subbands due to the five atomic layers. Each of the
subbands crosses the energy axis at the symmetry points with
vanishing slope. In other words, there are states with zero group
velocity which are responsible for the singularities in the LDOS
of each layer and will be shown below in the case of $x=0$ and
$x=1$. Since the behavior of electronic states and the
modification of energy band due to the influence of chemical
disorder play a significant role in determining the optical
properties of the disordered system, the LDOS should be analyzed
in detail.

In Figs. 2 and 3 we plot the layer-dependence of LDOS and the
optical absorption spectrum for $\Delta=0.5$ which corresponds to
$E_B=-3\,t$. Because of the symmetry of the system in the
$z$-direction, the electronic states for the layers $n$=1 and
$n=5$, and also for $n=2$ and $n=4$ are similar to each other. For
this reason, we have investigated our desired quantities in the
layers $n$=1, 2, and 3. The figures show how the value of alloy
concentration influences the $g_n(\omega)$ and $A_n(\omega)$. In
the case of $x=0$ and $x=1$, the LDOS of all layers shows a
steplike behavior for the states around the bottom and the top of
the band, while, sharp features are observed for energies around
the center of band. Both the steplike and sharp features (van-Hove
singularities) in the LDOS, which correspond to the symmetry
points of the 1BZ, comes from the two-dimensional nature of the
multilayer. The position of sharp peaks in the electronic density
of states is different in various layers due to the quantum
confinement along the confined direction. With increasing $x$ the
electronic spectra are shifted to the low-energy side and the
steplike behavior and sharp peaks gradually disappeared. In the
case of maximum disorder, i.e., at $x=0.5$ the LDOS is completely
symmetric with respect to the center of band. The steplike and
sharp features in the localized states are reproduced for $x>0.5$.
Also, it is clear that, a cusp may appear in the LDOS, depending
on the value of alloy concentration. This effect, which is related
to the value of $\Delta$, confirms that such electronic spectra
belong to the amalgamation type.
\begin{figure}
\centerline{\includegraphics[width=1.1\linewidth]{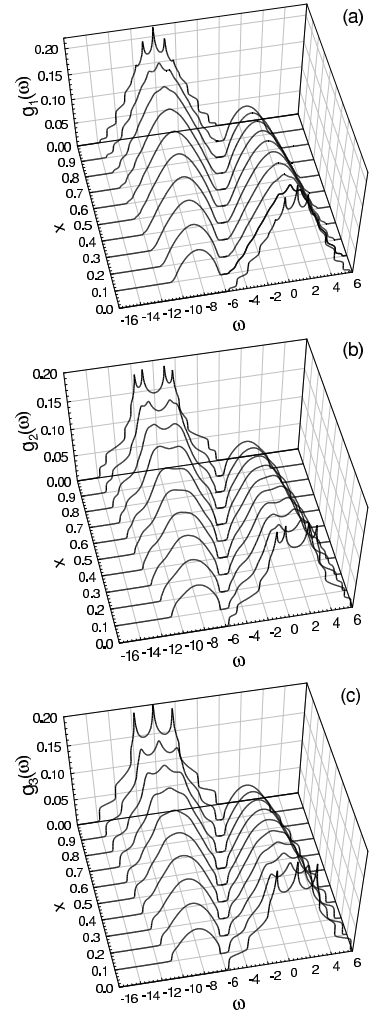}}
\caption{The same as Fig. 2 but for $E_B=-8\,t$.}
\end{figure}

\begin{figure}
\centerline{\includegraphics[width=1.1\linewidth]{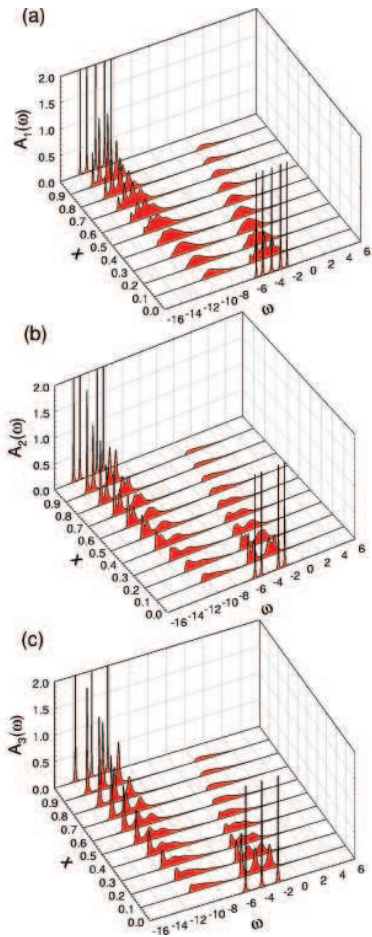}}
\caption{(Color online) The same as Fig. 3 but for $E_B=-8\,t$.}
\end{figure}

In a clean system, i.e., in the case of $x=0$ or $x=1$,
$A_n(\omega)$ shows sharp peaks at the edges of the subbands.
However, in a system with substitutional disorder, the absorption
peaks are broadened and their heights decrease. Based on the
behavior of electronic states, with increasing $x$, as shown in
Fig. 3, the optical absorption spectrum shifts to the low-energy
side, due to its dependence on the bottom energy of the subbands.
The maximum value of $A_n(\omega)$ corresponds to the cases of
$x=0$ and $x=1$, because the van Hove singularities for these
values of $x$ are significantly large. Thus, we can conclude that,
in doped semiconductor multilayers, impurities with lower (higher)
site energy in comparison with the energy of host atoms and with
finite concentration $x$, decrease the strength of optical
absorption, i.e., the height of peaks in $A_n(\omega)$, and
incident photons with low (high) energy are needed.
\begin{figure}
\centerline{\includegraphics[width=0.99\linewidth]{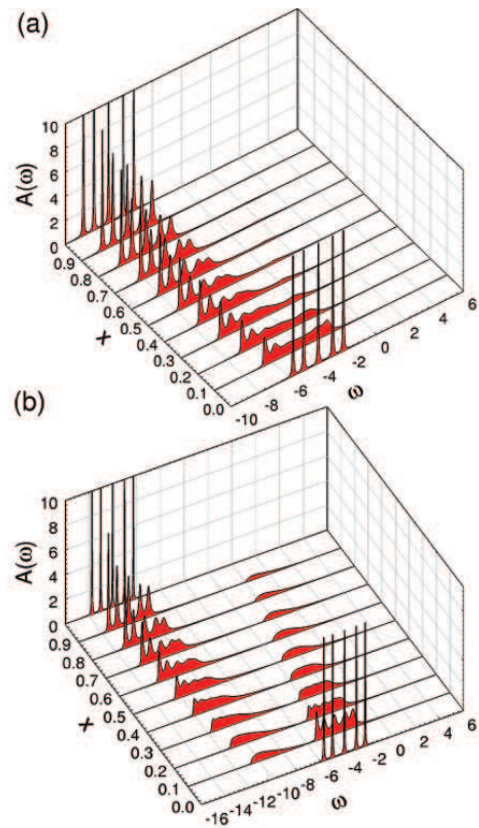}}
\caption{(Color online) The total optical absorption spectrum for
two disordered semiconductor multilayers as a function of energy
and alloy concentration. In (a) $E_B=-3\,t$ and in (b)
$E_B=-8\,t$. $\omega$ is measured in units of $t$.}
\end{figure}

Another interesting feature is that optical absorption of the
surface layers, i.e., layers $n=1$ and $n=N$, is smaller than that
of the interior layers, due to the influence of surface states
[compare the maximum value of $A_1(\omega)$ with that of
$A_2(\omega)$ and $A_3(\omega)$ at the lowest energies in Fig. 3].
This indicates that in multilayer systems, the value of optical
absorption in each layer depends on the layer number, and the
interior layers play a major role in such a process. This feature,
which has not been predicted in the previous studies
\cite{Shinozuka1,Shinozuka2}, is important in operation of
semiconductor optoelectronic nanodevices.

To investigate the another aspect of the electronic and optical
properties of the system, we have shown in Figs. 4 and 5,
$g_n(\omega)$ and $A_n(\omega)$ for $\Delta=1.33$ which
corresponds to $E_B=-8\,t$. We can clearly see that a gap appears
in the LDOS for each value of $x\,(\neq 0,1)$ and the electronic
spectra split into two bands corresponding to the two constituent
crystals; the lower- (higher-) energy band in this figure
corresponds to the $B$ ($A$) atoms. Accordingly, these electronic
spectra belong to the persistence type. This gap opening is a
consequence of scattering process of carriers by random
distribution of the elements in the alloy, and its magnitude
depends on the strength of disorder. Our analysis of the behavior
of LDOS is similar to that of Fig. 2.

Fig. 5 clearly indicates that in the case of strong disorder
regime, $A_n(\omega)$ consists of two bands and the spectrum
stretches toward the higher-energy side. With increasing $x$, the
optical absorption spectrum of the $A$-type band decreases and
shifts to the high-energy side, while the spectrum of the $B$-type
band increases and shifts to the low-energy side. Here, we should
mention that, under certain conditions, the width of optical
absorption spectrum in disordered quantum wires can be
considerably stronger than that of the present system
\cite{Fuchs}. An interesting point in Fig. 5 is that, in spite of
the symmetry of $g_n(\omega)$ for $x=0.5$ with respect to the
center of electronic spectrum, the optical spectrum is not
symmetric and the lower band shows a higher value of absorption
which is due to the fact that only the $\mathbf{k}_\parallel=0$
component contributes to this process.

It is important to mention that the optical absorption spectrum
from each layer is not a measurable quantity in experiment and
only the bulk (total) optical absorption spectrum, $A(\omega)$, is
possible. In order to obtain this quantity we should sum over the
optical absorption of all layers, i.e.,
$A(\omega)=\sum_{n=1}^{N_z}A_n(\omega)$. The total optical
absorption are shown in Fig. 6(a) and 6(b) for $E_B=-3\,t$ and
$E_B=-8\,t$, respectively. From the figures we see that, in the
case of $E_B=-3\,t$, the absorption spectrum has sharp peaks at
the bottom of the optical band for all values of $x$, and shows
that the value of total optical absorption at the band minimum of
the amalgamation type (Fig. 6(a)) is higher than that of the
persistence type (Fig. 6(b)). The numerical results also confirm
that, as in the bulk materials, the transition from the
amalgamation type to the persistence type in multilayers depends
on two parameters $\Delta$ and $x$ \cite{Onodera,Toyozawa}.

\section{Conclusion}
Using the single-band tight-binding theory and the single-site
CPA, we have studied the influence of substitutional disorder on
the electronic and optical properties of disordered semiconductor
multilayers. The results of this theory which is able to predict
either the persistent or amalgamated bands, indicate that the
optical absorption is governed by the electronic density of
states, which is stronger near the bottom of the band. The values
of absorption depend on the layer number, the concentration of
chemical disorder, and the scattering-strength parameter. In
addition, we found that the interior layers in comparison with the
surface layer, have significant contribution in the optical
absorption of the system.

The obtained results clearly indicate that, studying the
dependence of quantum size effects on the optical properties of
doped semiconductors is important for better understanding the
process of optical absorption in layered structures, and will be
helpful for designing optoelectronic nanodevices.

\end{document}